\documentclass[reprint,amsmath,amssymb,superscriptaddress,aps,prc,groupedaddress]{revtex4-1}

\usepackage[utf8]{inputenc}
\usepackage[english]{babel}
\usepackage[dvips]{graphicx}
\usepackage{epsfig}
\usepackage{amsmath}
\usepackage[dvips]{color}
\usepackage{graphics}
\usepackage{verbatim}
\usepackage{amssymb}
\usepackage{array}

\newcommand{\csi}{\textquotedblleft}

\begin{document}
\title{Precise determination of the effective-range parameters up to an arbitrary order}
\author{O. L. Ramírez Su\'arez}
\author{J-M. Sparenberg}
\affiliation{Physique Nucléaire et Physique Quantique, C.P. 229, Université Libre de Bruxelles (ULB), B 1050 Brussels, Belgium}
\date{\today}

\begin{abstract}
We propose a method to compute, for a given potential model, an arbitrary coefficient of the effective-range function expanded as a power series in energy. The method is based on a set of recurrence relations at low energy, that allows a compact and general description to any order in energy for neutral and charged cases. By using the Lagrange mesh technique to compute the R-matrix at zero energy, this proposal permits to compute, with a very good precision, the effective-range parameters. We use a potential model for some nuclear systems to illustrate the effectiveness of this method and to discuss its numerical limitations.
\end{abstract}

\maketitle

\section{Introduction}
Since the early days of nuclear physics, the study of scattering processes at low energies has been based on the partial-wave analysis of cross sections in terms of phase-shifts $\delta_l$, where $l$ is the orbital quantum number. Among the first systems studied, the description of the $s$-wave elastic scattering for the neutron-proton system took a high interest \cite{PR26,PR49}. For this system in particular, a good approximation of the function $k\cot(\delta_0)$ (where $k$ is the wave number in the center-of-mass frame) as a linear function of $k^2$ led to study the Effective-Range Function (ERF) and its expansion in Maclaurin series, better known as the Effective-Range Expansion (ERE). Since then, an ERF has been established for arbitrary partial waves, both in the neutral and charged cases \cite{PR83}. The analyticity properties of this function have also been demonstrated, making of EREs an important tool for the parametrization of low-energy scattering data.

The coefficients of the ERE, which are directly related to the so-called effective-range parameters, have become a field of study in themselves \cite{Newton,Bertulani,Goldberger}. These parameters, including the well-known scattering length, effective range and shape parameter, can be interpreted in terms of physical properties of the system, in particular regarding scattering states at low energies and weakly bound states \cite{PRC81}. They can be considered as parameters fitting these properties, hence providing a useful analysis tool for experimental data. For instance, Ref.\ \cite{PR49} shows a way to obtain the first parameters of the ERE for the neutron-proton and proton-proton systems from experiment. Of course, in many cases, the possibility of determining the effective-range parameters via the experimental data is open but the precision will depend on how close we can go to zero energy and how many parameters we wish to determine.

The ERE coefficients actually carry information about the interaction between the colliding particles; they can thus be considered as an intermediate step between experiment and models. In an inverse-problem perspective, they can be extracted from scattering data and then used as an input for an inversion method aimed at deriving an interaction model. Conversely, ERE coefficients are calculated for a given interaction by solving the Schrödinger equation at low-energy scattering, hence providing an efficient way of comparing the predictions from this model with experimental data. A efficient method to obtain the first three effective-range parameters is proposed in Refs.\ \cite{PRC61,PRC63}. In particular, the latter shows that using the R-matrix \cite{Burke,RPP73} and the Lagrange mesh technique \cite{PSS243,PRE84}, the prediction of the scattering length, the effective range and the shape parameters is accurate for neutral and charged cases. The aim of the present paper is to extend that proposal to calculate any parameter of the ERE in both cases, for a given interaction model.

In Sec.\ \ref{sec-ERE}, we introduce the ERF of a two-body system via the phase shift both for the neutral and charged case. Then, we define the effective-range parameters and present the method to compute any of them. In Sec.\ \ref{sec-LM} we briefly describe the scheme to compute derivatives at zero energy of the R-matrix based on the Lagrange mesh technique. In order to test the method, three neutral and three charged cases are analyzed in Sec.\ \ref{sec-app}. Conclusions and perspectives are presented in Sec.\ \ref{sec-CP}.

\section{Effective-range function at low energies}\label{sec-ERE}
Let us consider the scattering of two particles with relative coordinate $r$ and reduced mass $\mu$. Let us also suppose that part of the interaction between the two particles decreases fast enough in such a way that the effective potential can be described in two regions as the contribution of: a short-range potential ($V_\text{N}$), a centrifugal potential and a Coulomb potential ($V_\text{C}$) for $r \le a$ and a centrifugal and a Coulomb potentials for $r>a$. Here $a$ is considered as the range of the short-range potential but in practice it could be taken larger or equal to the minimal distance between the two particles when the short-range potential is neglected in comparison with the rest of the interaction. In this scenario, for the $l$th partial wave, and if $\Psi_l(k,r)=\frac{u_l(k,r)}{kr}$ is the radial wave function, the radial Schr\"odinger equation can be written as
\begin{equation}\label{SE}
\left[-\frac{\hbar^2}{2\mu}\frac{d^2}{dr^2}+\frac{\hbar^2}{2\mu}\frac{l(l+1)}{r^2}+V(r)-E\right]u_l(k,r)=0,
\end{equation}
where $u_l(k,r)$ is known as the modified radial wave function, $E=\frac{\hbar^2}{2\mu}k^2$ is the energy in the center of mass frame and
\begin{equation}\label{V}
V(r)=
\begin{cases}
V_\text{N}(r)+V_\text{C}(r), & \text{if } r \le a,\\
V_\text{C}(r), & \text{if } r>a.
\end{cases}
\end{equation}

Let us introduce the following notations and conventions:
\begin{list}{$\bullet$}{\setlength{\leftmargin}{0.6cm}\setlength{\rightmargin}{0cm}}
\item ${H_1}_l={H_1}_l(k,a)$ is the Ricatti-Bessel (regular Coulomb) function at $r=a$ for the neutral (charged) case.
\item ${H_2}_l={H_2}_l(k,a)$ is the Ricatti-Neumann (irregular Coulomb) function at $r=a$ for the neutral (charged) case.
\item Prime represents the partial derivative with respect to $r$ at $r=a$ (e.g.,  ${H'_i}_l=\left.\partial{H_i}_l(k,r)/\partial r\right|_{r=a}$).
\item $R_l=\left[a \left.\frac{\partial \ln(u_l(k,r))}{\partial r}\right|_{r=a}\right]^{-1}$ is the single channel R-matrix \cite{Burke,RPP73}.
\item For the charged case, if the interacting particles have electric charge $Z_1$e and $Z_2$e, then $a_N=\frac{\hbar^2}{\mu}\frac{4\pi\epsilon_0}{Z_1 Z_2 \text{e}^2}$ is the nuclear Bohr radius, $E_N=\frac{\hbar^2}{2\mu}\frac{1}{a_N^2}$ is the nuclear Rydberg energy and 
\begin{equation}\label{eta}
\eta=\frac{1}{a_N k}=\sqrt{\frac{E_N}{E}}
\end{equation}
is the Sommerfeld parameter.
\end{list}
With the previous notations and conventions, the phase-shift $\delta_l$ \cite{Bertulani} is obtained through
\begin{equation}\label{psl}
\tan(\delta_l)=-\frac{{H_1}_l-aR_l{H'_1}_l}{{H_2}_l-aR_l{H'_2}_l}.
\end{equation}

Equation \eqref{psl} gives a good tool to compute the phase-shifts and later the ERF in a wide range of energy. Unfortunately Eq.\ \eqref{psl} is not useful at low energies, in particular at the limit when $E$ tends towards zero, because the functions ${H_1}_l$ and ${H_2}_l$ and their radial derivatives vanish, giving an indeterminate value for the phase-shifts and then for the ERF. However, this limit is very important to determine the coefficients of the ERE, since by definition they are calculated at zero energy.

In order to overcome the indeterminacy of $\delta_l$ at $E=0$, a convenient re-normalization of the ${H_1}_l$ and ${H_2}_l$ functions is proposed in Ref.\ \cite{PRC63}. These renormalized functions remain finite at zero energy and are expressed as
\begin{align}
{\mathcal{H}_1}_l&=
\begin{cases}
k^{-l-1}{H_1}_l, 				&\text{(neutral case)},\\
k^{-1/2}\exp(\pi\eta){H_1}_l,	&\text{(charged case)},
\end{cases}\label{H1}\\
{\mathcal{H}_2}_l&=
\begin{cases}
k^{l}{H_2}_l,								&\text{(neutral case)},\\
\frac{\pi}{2}k^{-1/2}\exp(-\pi\eta){H_2}_l,	&\text{(charged case)}.
\end{cases}\label{H2}
\end{align}
Using definitions \eqref{H1} and \eqref{H2} in Eq.\ \eqref{psl} one can easily obtain the expressions for the phase-shits
\begin{equation}\label{psl2}\\
\tan(\delta_l)=
\begin{cases}
k^{2l+1}D_l, 						&\text{(neutral case)},\\
\frac{\pi}{2}\exp(-2\pi\eta)D_l,	&\text{(charged case)},
\end{cases}
\end{equation}
with
\begin{equation}\label{Dl}
D_l=-\frac{{\mathcal{H}_1}_l-aR_l{\mathcal{H}'_1}_l}{{\mathcal{H}_2}_l-aR_l{\mathcal{H}'_2}_l}.
\end{equation}
For simplicity, we have omitted the dependence on energy in Eqs.\ \eqref{psl2} and \eqref{Dl}. 

By using the previous notation the ERF, $K_l$, can be written as \cite{PR83}
\begin{equation}\label{Kl}
K_l=
\begin{cases}
D_l^{-1}, 					&\text{(neutral case)},\\
\frac{2w_l}{l!^2a_N^{2l+1}}\left[2D_l^{-1}+h\right],	&\text{(charged case)},
\end{cases}
\end{equation}
where the functions $w_l$ and $h$ read
\begin{gather}
w_l=\prod_{j=0}^l\left[1+\frac{j^2}{\eta^2}\right],\label{wl}\\
h=\psi(i\eta)-\ln(i\eta)+\frac{1}{2i\eta}-\frac{i\pi}{e^{2\pi\eta}-1},\label{h}
\end{gather}
with $\psi$ the digamma function \cite{Abramowitz}.

It can be proven that $K_l$ is analytical at zero energy, which allows to write this function as a power series in energy, i.e.,
\begin{equation}\label{ERE2}
K_l=\sum_{n=0}^\infty c_{l,n} E^n.
\end{equation}

Equation \eqref{ERE2} is useful to define the effective-range parameters. Here we adopt the standard convention for the first three efective-range parameters: the scattering length ($a_l$), the effective range ($r_l$) and the shape parameter ($P_l$), i.e., expanding Eq.\ \eqref{ERE2} one finds 
\begin{align}
K_l=
&-\frac{1}{a_l}+\frac{r_l}{2}\left(\frac{2\mu}{\hbar^2}\right)E-P_lr_l^3\left(\frac{2\mu}{\hbar^2}\right)^2E^2\notag\\
&+\sum_{n=3}^\infty Q_{l,n}\left(\frac{2\mu}{\hbar^2}\right)^n E^n.\label{ERE3}
\end{align}

In Eq.\ \eqref{ERE3} we have defined $Q_{l,n}=\left(\frac{\hbar^2}{2\mu}\right)^nc_{l,n}$ as the effective-range parameter of order $n$, for $n\geq 3$.

Our goal is to compute the coefficient $c_{l,n}$ in Eq.\ \eqref{ERE2} and then to calculate the effective-range parameters in Eq.\ \eqref{ERE3}. The coefficient $c_{l,n}$ is easily found by computing the $n$th derivative of $K_l$ at zero energy
\begin{equation}\label{cn}
c_{l,n}=\frac{1}{n!}{(K_l)}_0^{(n)}.
\end{equation}
Hereafter, \textit{we use the superscript $(n)$ to designate the $n$th partial derivative with respect to energy and the subscript $0$ to designate functions at $E=0$}. The reader should not confuse the prime with the superscript $(1)$.

As we have shown in Eq.\ \eqref{Kl}, the explicit form of $K_l$ depends on whether the Coulomb interaction is present in the system or not. For this reason we will study both cases separately in the next two subsections. 

\subsection{$c_{l,n}$ for the neutral case}
In the neutral case $c_{l,n}$ depends on $D_l^{-1}$ only, then using Eq.\ \eqref{Dl} one can rewrite Eq.\ \eqref{cn} as
\begin{equation}\label{cnn1}
c_{l,n}=\frac{1}{n!}\underbrace{\left(-\frac{{\mathcal{H}_2}_l-aR_l{\mathcal{H}'_2}_l}{{\mathcal{H}_1}_l-aR_l{\mathcal{H}'_1}_l}\right)^{(n)}_0}_{\left(D_l^{-1}\right)^{(n)}_0}.
\end{equation}
The term $(D_l^{-1})^{(n)}_0$ can be expanded in terms of the derivatives of ${\mathcal{H}_i}_l$, ${\mathcal{H}'_i}_l$ and $R_l$ at $E=0$. To do this, let us first define $D_l^{-1}$ as the following
\begin{equation}\label{Dlin}
D_l^{-1}=-\frac{{\Delta_2}_l}{{\Delta_1}_l},
\end{equation}
with
\begin{equation}
{\Delta_i}_l={\mathcal{H}_i}_l-aR_l{\mathcal{H}'_i}_l, \quad \text{for } i\in\{1,2\}.
\end{equation}
Using the product rule for derivatives, one gets
\begin{equation}\label{dDlinv}
\left(D_l^{-1}\right)^{(n)}_0=-\sum_{m=0}^n\binom{n}{m}\left({\Delta_2}_l\right)_0^{(n-m)}\left({\Delta_1}_l^{-1}\right)_0^{(m)},
\end{equation}
with $\binom{n}{m}$ the binomial coefficient. In the same way, the terms $\left({\Delta_i}_l\right)_0^{(n)}$ and $\left({\Delta_i}_l^{-1}\right)_0^{(n)}$ can be written as
\begin{equation}\label{dDelta}
\left({\Delta_i}_l\right)^{(n)}_0=\left({\mathcal{H}_i}_l\right)_0^{(n)}-a\sum_{j=0}^n\binom{n}{j}(R_l)^{(n-j)}_0({\mathcal{H}'_i}_l)^{(j)}_0
\end{equation}
and
\begin{multline}\label{dDeltainv}
\left({\Delta_i}_l^{-1}\right)^{(n)}_0=-\sum_{m=1}^n\sum_{j=0}^{n-m} 
\binom{n-1}{m-1} \binom{n-m}{j} \left({\Delta_i}_l\right)^{(m)}_0\\
\times
\left({\Delta_i}_l^{-1}\right)_0^{(n-m-j)}
\left({\Delta_i}_l^{-1}\right)_0^{(j)},
\end{multline}
for $n>0$.

The four previous equations, together with Eq.\ \eqref{cnn1}, allow us to compute recursively any coefficient $c_{l,n}$ if one knows $({\mathcal{H}_i}_l)^{(n)}_0$, $({\mathcal{H}'_i}_l)^{(n)}_0$ and $(R_l)^{(n)}_0$. The value of $({\mathcal{H}'_i}_l)^{(n)}_0$ is easily obtained from $({\mathcal{H}_i}_l)^{(n)}_0$ since the derivatives over $r$ and $E$ can be interchanged. The functions $({\mathcal{H}_1}_l)^{(n)}_0$ and $({\mathcal{H}_2}_l)^{(n)}_0$ read \cite{PRC63}
\begin{equation}\label{mathH1}
({\mathcal{H}_1}_l)^{(n)}_0=\left(-\frac{\mu}{\hbar^2}\right)^n\frac{a^{l+2n+1}}{(2l+2n+1)!!},
\end{equation}
\begin{equation}\label{mathH2}
({\mathcal{H}_2}_l)^{(n)}_0=\left(\frac{\mu}{\hbar^2}\right)^n(2l-2n-1)!!a^{-l+2n}.
\end{equation}
Thus, there only remains to find a general expression for $(R_l)^{(n)}_0$ to complete all requirements to compute the coefficient $c_{l,n}$. An efficient method to get the derivatives of the R-matrix at zero energy is described in Ref.\ \cite{PRC63}. We will briefly explain it in Sec.\ \ref{sec-LM}.

\subsection{$c_{l,n}$ for the charged case}\label{sec-met-charged}
By using Eqs. \eqref{Kl} and \eqref{cn}, the coefficient $c_{l,n}$ for the charged case reads
\begin{equation}\label{cnch1}
c_{l,n}=\frac{1}{n!}\frac{2}{l!^2a_N^{2l+1}} \left[2\left(w_lD_l^{-1}\right)_0^{(n)}+(w_lh)_0^{(n)}\right].
\end{equation}
Note that one can use the same procedure as the one used for the neutral case to split the derivatives of the products $w_lD_l^{-1}$ and $w_lh$. With this idea the term $\left(w_lD_l^{-1}\right)_0^{(n)}$ can be written as
\begin{equation}\label{dwlDlinv}
\left(w_lD_l^{-1}\right)_0^{(n)}=\sum_{m=0}^n\binom{n}{m}(w_l)_0^{(n-m)} \left(D_l^{-1}\right)_0^{(m)}
\end{equation}
and using the power series expansion of the function $h$
\begin{align}
\hspace{-0.08cm}h=\sum_{n=0}^\infty\tilde{h}_n E^n=
&\frac{1}{12}\frac{E}{E_N}+\frac{1}{120}\frac{E^2}{E_N^2}+\frac{1}{252}\frac{E^3}{E_N^3}+\frac{1}{240}\frac{E^4}{E_N^4}\notag\\
&+\frac{1}{132}\frac{E^5}{E_N^5}+\frac{691}{32760}\frac{E^6}{E_N^6}+\frac{1}{12}\frac{E^7}{E_N^7}\notag\\
&+\frac{3617}{8160}\frac{E^8}{E_N^8}+\frac{43867}{14364}\frac{E^9}{E_N^9}+\cdots,\label{hserie1}
\end{align}
the term $(w_lh)_0^{(n)}$ reads
\begin{equation}\label{dwlh}
(w_lh)_0^{(n)}=\sum_{m=0}^n\frac{n!}{(n-m)!}(w_l)_0^{(n-m)} \tilde{h}_m.
\end{equation}

The term $\left(D_l^{-1}\right)_0^{(m)}$ in Eq.\ \eqref{dwlDlinv} can be computed by using Eqs.\ \eqref{dDlinv}-\eqref{dDeltainv}. For $n>0$ the term $(w_l)_0^{(n)}$ reads
\begin{equation}\label{dwl}
(w_l)_0^{(n)}=\sum_{m=0}^{n-1}(-1)^m\frac{(n-1)!}{(n-1-m)!}\frac{(w_l)_0^{(n-1-m)}}{E_N^{m+1}}N_{l,m},
\end{equation}
with 
\begin{equation}\label{Nlk}
N_{l,m}=\sum_{j=1}^l j^{2(m+1)}.
\end{equation}
The value $(w_l)_0^{(0)}=1$ is easily calculated from Eq.\ \eqref{wl}. 

In Eq.\ \eqref{hserie1} we have written the first ten terms explicitly, which will be used in Sec.\ \ref{sec-app} to compute the first ten effective-range parameters. Note that $\tilde{h}_0=0$.

At this point we have all the requirements to compute $c_{l,n}$ except for general expressions of $(R_l)^{(n)}_0$, $({\mathcal{H}'_i}_l)^{(n)}_0$ and $({\mathcal{H}_i}_l)^{(n)}_0$ which are required to calculate Eq.\ \eqref{dDelta}. Similarly to the neutral case, the term $(R_l)^{(n)}_0$  will be obtained by using the Lagrange mesh technique described in Sec.\ \ref{sec-LM}, and the term $({\mathcal{H}'_i}_l)^{(n)}_0$ will be easily computed if one knows  $({\mathcal{H}_i}_l)^{(n)}_0$. Therefore, we can focus on finding a general expression for $({\mathcal{H}_i}_l)^{(n)}_0$.

Let us start with the regular Coulomb function ${H_1}_l$. For $\eta > 0$ this function can be expanded as
\begin{equation}\label{H1serie}
{H_1}_l=\sqrt{\frac{\pi}{2\eta}\frac{w_l}{e^{2\pi\eta}-1}}
\left(\frac{2}{x}\right)^{2l}\sum_{m=2l+1}^\infty b_m\left(\frac{x}{2}\right)^m I_m(x),
\end{equation}
where we have started from its expansion in terms of Bessel-Clifford functions \cite{Abramowitz}. The argument of the first modified Bessel function $I_m$ is the dimensionless variable $x=2\sqrt{\frac{2r}{a_N}}$. The term $b_m$ satisfies the recurrence relation
\begin{equation}\label{recbk}
4\eta^2(m-2l)b_{m+1}+mb_{m-1}+b_{m-2}=0,
\end{equation}
for $m>2l+1$ and with $b_{2l}=b_{2l+2}=0$, $b_{2l+1}=1$.

By using Eqs.\ \eqref{H1serie} in Eq.\ \eqref{H1} one finds the compact expression for the renormalized Coulomb function
\begin{equation}\label{rH1serie}
{\mathcal{H}_1}_l=\sqrt{\pi r}\sum_{m=2l+1}^\infty d_m\phi_m(x,l),
\end{equation}
where the terms $d_m=d_m(\eta,l)$ and $\phi_m(x,l)$ read
\begin{equation}\label{dk}
d_m=b_m\sqrt{\frac{w_l}{1-e^{-2\pi\eta}}}
\end{equation}
and
\begin{equation}\label{phik}
\phi_m(x,l)=\left(\frac{x}{2}\right)^{m-(2l+1)}I_m(x).
\end{equation}
 
Because of the fast decrease of the term $e^{-2\pi\eta}$ when the energy goes to zero, one can check from Eq.\ \eqref{dk} that the only contribution to the derivatives of $d_m$ at zero energy is given by the term $b_m\sqrt{w_l}$. Therefore, from Eq.\ \eqref{rH1serie} one obtains 

\begin{equation}\label{dErH1serie}
({\mathcal{H}_1}_l)_0^{(n)}=\sqrt{\pi r}\sum_{m=2l+1}^\infty (b_m\sqrt{w_l})_0^{(n)}\phi_m(x,l).
\end{equation}

Using the product rule for derivatives and taking into account that $b_m$ is a polynomial in energy, Eq.\ \eqref{dErH1serie} can be written as
\begin{multline}\label{dErH1serie2}
({\mathcal{H}_1}_l)_0^{(n)}=\sqrt{\pi r}\sum_{j=0}^n\sum_{m=2(n-j)}^{3(n-j)}\Biggl[\binom{n}{j}\left(\sqrt{w_l}\right)_0^{(j)} \\
\times\left(b_{m+2l+1}\right)_0^{(n-j)}\left(\frac{x}{2}\right)^{m}I_{m+2l+1}(x)\Biggr].
\end{multline}
Note that the sum over $m$ runs from $2(n-j)$ up to $3(n-j)$. This is because $\left(b_{m+2l+1}\right)_0^{(\bar{m})}=0$ if $m<2\bar{m}$ or $m>3\bar{m}$, as one can prove from Eq.\ \eqref{recbk}.

The terms $\left(\sqrt{w_l}\right)_0^{(n)}$ and $(b_{m+1})_0^{(n)}$ can be calculated by iterating 
\begin{equation}\label{dsqrtwl}
\left(\sqrt{w_l}\right)_0^{(n)}=\frac{(n-1)!}{2E_N^n}\sum_{j=0}^{n-1}(-1)^j\frac{\left(\sqrt{w_l}\right)_0^{(n-1-j)}}{(n-1-j)!}N_{l,j},
\end{equation}
with $N_{l,j}$ defined in Eq.\ \eqref{Nlk}, and
\begin{equation}\label{dbk}
(b_{m+1})_0^{(n)}=-n\frac{m(b_{m-1})_0^{(n-1)}+(b_{m-2})_0^{(n-1)}}{4(m-2l)E_N},
\end{equation}
for $m>1$ and $n>0$.

For $i\in\{1,2\}$, the simple and useful relation
\begin{equation}\label{ddHi}
({\mathcal{H}'_i}_l)^{(n)}_0=\frac{1}{2a} \left[({\mathcal{H}_i}_l)^{(n)}_0+x\frac{\partial({\mathcal{H}_i}_l)^{(n)}_0}{\partial x}\right]_{r=a},
\end{equation}
helps us to compute $({\mathcal{H'}_i}_l)_0^{(n)}$ easily. 

If $i=1$ and using Eq.\ \eqref{dErH1serie2} one can compute the term $\frac{\partial({\mathcal{H}_i}_l)^{(n)}_0}{\partial x}$ in Eq.\ \eqref{ddHi} as
\begin{widetext}
\begin{equation}\label{ddH1}
\frac{\partial({\mathcal{H}_1}_l)^{(n)}_0}{\partial x}=
\sqrt{\pi r} \sum_{j=0}^n \sum_{m=2(n-j)}^{3(n-j)}
\left[
\binom{n}{j} \left(\sqrt{w_l}\right)_0^{(j)} 
\left(b_{m+2l+1}\right)_0^{(n-j)} \left(\frac{x}{2}\right)^m
\left(\frac{m}{x}I_{m+2l+1}(x) + \frac{dI_{m+2l+1}(x)}{dx}\right)
\right].
\end{equation}
\end{widetext}

We have discussed the steps to obtain $({\mathcal{H}_i}_l)^{(n)}_0$ and $({\mathcal{H}'_i}_l)^{(n)}_0$ if $i=1$. We can now move on to the case $i=2$. For this case, it is necessary to clarify that the expansion in Bessel-Clifford functions of the irregular Coulomb wave function ${H_2}_l$ \cite{Abramowitz} is possible only when $\eta\to\infty$. This limit is equivalent to $E\to 0$ (see Eq.\ \eqref{eta}) which is our case of interest. Therefore, mimicking the procedure followed previously for $({\mathcal{H}_1}_l)^{(n)}_0$ and  $({\mathcal{H}'_1}_l)^{(n)}_0$, and defining $\bar{I}_n$ as the second modified Bessel function, we find
\begin{widetext}
\begin{gather}
({\mathcal{H}_2}_l)_0^{(n)}=\sqrt{\pi r}\sum_{j=0}^n\sum_{m=2(n-j)}^{3(n-j)}\left[(-1)^m\binom{n}{j}\left(\sqrt{w_l}\right)_0^{(j)}
\left(b_{m+2l+1}\right)_0^{(n-j)} \left(\frac{x}{2}\right)^{m}\bar{I}_{m+2l+1}(x)\right],\label{dErH2serie2}\\
\frac{\partial({\mathcal{H}_2}_l)^{(n)}_0}{\partial x}=
\sqrt{\pi r} \sum_{j=0}^n \sum_{m=2(n-j)}^{3(n-j)}
\left[(-1)^m
\binom{n}{j} \left(\sqrt{w_l}\right)_0^{(j)} 
\left(b_{m+2l+1}\right)_0^{(n-j)} \left(\frac{x}{2}\right)^m
\left(\frac{m}{x}\bar{I}_{m+2l+1}(x) + \frac{d\bar{I}_{m+2l+1}(x)}{dx}\right)
\right].\label{ddH2}
\end{gather}
\end{widetext}

\section{Lagrange mesh technique to compute R-matrix derivatives at zero energy}\label{sec-LM}
In order to understand some properties of two-body nuclear systems, it is convenient to take advantage of the fact that the nuclear interaction is a short-range interaction. This fact allows us to divide the configuration space in two regions as it is shown in Eqs.\ \eqref{SE} and \eqref{V}. In the external region ($r>a$) the Schr\"odinger equation is solved by analytic procedures. The challenge is to find the wave function in the internal region ($r<a$) together with the correct matching between the external and internal wave functions at $r=a$. It is equivalent to find the R-matrix for a single-channel case. In general, using analytic methods it is not possible to achieve this challenge (at least nowadays) and for this reason one needs to find a good approximation of the wave function, which is usually done numerically. In the case of the R-matrix $R_l$, Ref.\ \cite{RPP73} shows how to achieve this challenge by approximating the wave function as a superposition of $N$ linearly independent functions, $f_n(r)$, of a basis. The result for a two-body system with reduced mass $\mu$ and energy $E$ is
\begin{equation}\label{Rmat}
R_l\approx\frac{\hbar^2}{2\mu a^2}\sum_{n,m=1}^N f_n(a)[(C-EI)^{-1}]_{nm}f_m(a),
\end{equation}
where the elements of the $N\times N$ matrix $I$ and $C$ are respectively the Kronecker delta $\delta_{nm}$ and
\begin{equation}\label{Cnmint}
C_{nm}=\int_0^a f_n(r)\left[T_l+V(r)+\mathcal{L}\right]f_m(r)dr,
\end{equation}
with
\begin{equation}
T_l=-\frac{\hbar^2}{2\mu}\left[\frac{d^2}{dr^2}-\frac{l(l+1)}{r^2}\right],
\end{equation}
the kinetic energy operator of the partial wave $l$, and
\begin{equation}
\mathcal{L}=\frac{\hbar^2}{2\mu}\delta(r-a)\frac{d}{dr},
\end{equation}
the Bloch operator \cite{NP4}. The latter appears for two main reasons. First, ``to keep" the hamiltonian hermitian in the internal region and second, to match the internal and external wave functions correctly.

The good estimation of $C_{nm}$ in Eq.\ \eqref{Cnmint} will depend on the basis chosen and how the integral is made. As the internal space is $r\in[0,a]$, a good basis is the shifted Lagrange-Legendre basis \cite{RPP73}. This basis allows to compute the integral in Eq.\ \eqref{Cnmint} easily if it is approximated by the shifted Gauss-Legendre quadrature. With these assumptions $f_n(a)$ and $C_{nm}$ read \cite{PRC63,RPP73}
\begin{gather}
f_n(a)=\frac{(-1)^n}{\sqrt{ax_n(1-x_n)}}, \quad \text{for } n\in\{1,2,\cdots,N\},\label{fn}\\
C_{nm}\approx
\begin{cases}
\frac{\hbar^2}{2\mu}\frac{(4N^2+4N+3)x_n(1-x_n)-6x_n+1}{3a^2x_n^2(1-x_n)^2}\\
+\frac{\hbar^2}{2\mu}\frac{l(l+1)}{a^2x_n^2}+V(ax_n), \qquad\text{ for } n=m,\\
\\
\frac{\hbar^2}{2\mu}\frac{(-1)^{n+m}}{a^2\sqrt{x_nx_m(1-x_n)(1-x_m)}}\\
\times\left[N^2+N+1+\frac{x_n+x_m-2x_nx_m}{(x_n-x_m)^2}\right.\\
\,\,\quad\left.-\frac{1}{1-x_n}-\frac{1}{1-x_m}\right], \qquad\text{ for } n\neq m,\\
\end{cases}\label{Cnm}
\end{gather}
with $x_n$ the abscissas of a shifted Gauss-Legendre quadrature in the [0,1] interval.

Finally, the previous approximations allow to calculate the $j$th derivative of the R-matrix at zero energy as
\begin{equation}\label{dRl0}
(R_l)_0^{(j)}\approx\frac{\hbar^2}{2\mu}\frac{j!}{a}\sum_{n,m=1}^Nf_n(a)[(C^{-1})^{j+1}]_{nm}f_m(a).
\end{equation}

\section{Applications}\label{sec-app}
In this section we discuss the effectiveness of the method derived in Sec.\ \ref{sec-ERE} to calculate the first ten effective-range parameters via the coefficients $c_{l,n}$ in Eqs.\ \eqref{cnn1} and \eqref{cnch1}. To do that, we use the technique shown in Sec.\ \ref{sec-LM} to compute the R-matrix derivatives at zero energy. We analyze the following nuclear systems:

\begin{list}{}{\setlength{\leftmargin}{0cm}\setlength{\rightmargin}{0cm}}
\item \emph{Neutral systems:}
\begin{list}{$\bullet$}{\setlength{\leftmargin}{0.4cm}\setlength{\rightmargin}{0cm}}
\item SW (Square Well): $s,p,d$-waves, $\frac{\hbar^2}{2\mu}=1$ MeV fm$^2$ and square well potential of width 3 fm and depth 10 MeV.
\item W-S (Woods-Saxon): $s,p,d$-waves, $\frac{\hbar^2}{2\mu}=1$ MeV fm$^2$ and Woods-Saxon potential $\frac{-{V}_0}{1+\exp\left[(r-R)/d\right]}$, with $R=3$ fm and $V_0=10[1+\exp(-R/d)]$ MeV. The diffuseness $d$ will be taken as a variable parameter. 
\item n-p (neutron-proton): $s$-wave and Bargmann potential \cite{Newton} $-8b\beta^2\frac{\exp(-2\beta r)}{[1+b\exp(-2\beta r)]^2}\frac{\hbar^2}{2\mu}$, with $b=\frac{\beta-\alpha}{\beta+\alpha}$, $\alpha=0.04$ fm$^{-1}$ and $\beta=0.81$ fm$^{-1}$ \cite{PR107}.
\end{list}
\item \emph{Charged systems:}
\begin{list}{$\bullet$}{\setlength{\leftmargin}{0.4cm}\setlength{\rightmargin}{0cm}}
\item $\alpha+^3$He: $s$-wave, point-sphere Coulomb potential with $R_C=3.248$ fm and gaussian potential $-67.67\exp\left[-r^2/(2.477\text{ fm})^2\right]$ MeV \cite{PRC61}.
\item $^{16}$O$+$p: $p$-wave, point-sphere Coulomb potential with $R_C=3.553$ fm and gaussian potential $-36\exp\left[-r^2/(3.553\text{ fm})^2\right]$ MeV \cite{PRC61}.
\item $^{12}$C$+\alpha$: $d$-wave, erf-Coulomb potential $\frac{Z_1Z_2e^2}{4\pi\epsilon_0r}\text{erf}\left(\frac{r}{2.5\text{ fm}}\right)$, where erf is the error function \cite{Abramowitz}, and gaussian potential $-112.3319\exp\left[-r^2/(2.8\text{ fm})^2\right]$ MeV \cite{PRC81}.
\end{list}
\end{list}

All the previous systems are chosen in order to explore the numerical merits and numerical limitations of our method as much as possible. We choose the SW and n-p systems because they allow us to compare our results with exact results. The W-S system is chosen to evaluate the behavior of the effective-range parameters when a weakly bound state, virtual state or resonance at low energy is present. The last three systems were chosen to evaluate the charged case and because, together with n-p, all of them are closer to real nuclear cases.

Before showing results, let us specify the calculation conditions. All masses are atomic masses and are taken from Ref.\ \cite{NPA729} except for systems SW and W-S. Units of the effective-range parameters and theirs absolute errors, $\varepsilon$, will be given in fm$^{\nu}$ where $\nu$ is shown in Table \ref{Tunits}.
\begin{table}[htb]
\caption{Units for the effective-range parameters and absolute error in fm$^{\nu}$.}
\label{Tunits}
\begin{ruledtabular}
\begin{tabular}{ccccc}
				& $a_l$, $\varepsilon(a_l)$	& $r_l$, $\varepsilon(r_l)$ 	& $P_l$, $\varepsilon(P_l)$ & $Q_{l,n}$, $\varepsilon(Q_{l,n})$	\\
\hline
$\nu$	& $2l+1$	& $-2l+1$ 	& $4l$ 	& $-2l+2n-1$
\end{tabular}
\end{ruledtabular}
\end{table}

In order to estimate the effective-range parameters using the method described in Sec.\ \ref{sec-ERE}, it is necessary to compute expressions that involve many operations that have to be carried out numerically. This fact leads to some numerical limitations in the accuracy of the estimations that will depend on the characteristics of the software and hardware used. Here we develop such a numerical evaluation using double precision in FORTRAN language on a regular desktop computer. With these conditions one expects that the values estimated match the exact result up to 15 digits in the best case.

\subsection{Neutral cases}\label{sec-app-neutral}
Exact values of the first ten effective range parameters for the SW and n-p systems are displayed in Table \ref{Texact}.
\begin{table}[htb]
\caption{Exact values of the first ten effective-range parameters for the SW and n-p systems. All values are written with twelve digits except $P_0$ and $Q_{0,n}$ for the n-p system. Units can be consulted in Table \ref{Tunits}.}
\label{Texact}
\begin{ruledtabular}
\begin{tabular}{*{4}{>{$}c<{$}}}
	& \multicolumn{3}{c}{SW} \\
\cline{2-4}
	& s$-wave$ & p$-wave$ & d$-wave$ \\
\hline
a_l		&  2.98035125196 &  54.5044450592 &  4.32199541189 \\
r_l		&  1.95321791835 & -1.01095733502 & -1.04292323145 \\ 
P_l		& -0.07224549975 &  0.69171350459 & -0.87150574727 \\ 
Q_{l,3}	&  0.41010095011 &  0.21472158944 &  0.35035876175 \\ 
Q_{l,4}	&  0.32190586543 &  0.10883657253 &  0.01058724882 \\ 
Q_{l,5}	&  0.25278100437 &  0.04415761104 & -0.01015831203 \\ 
Q_{l,6}	&  0.19808608111 &  0.02925502308 & -0.00625258947 \\ 
Q_{l,7}	&  0.15503303001 &  0.00791996459 & -0.00253456328 \\ 
Q_{l,8}	&  0.12127431544 &  0.00948504977 & -0.00082189556 \\ 
Q_{l,9}	&  0.09484976310 & -0.00012111901 & -0.00021419743 \\ 
\hline
	& \multicolumn{3}{c}{n-p} \\
\cline{2-4}
a_0		&  \multicolumn{3}{c}{$-23.7654320988$} \\
r_0		&  \multicolumn{3}{c}{2.59740259740} 	\\
P_0		&  \multicolumn{3}{c}{0} 				\\
Q_{0,n}	&  \multicolumn{3}{c}{0, for $n\geq3$}				
\end{tabular}
\end{ruledtabular}
\end{table}

Let us discuss some relevant features of the $Q_{l,n}$ parameters displayed in Table \ref{Texact} from the mathematical and numerical point of view. 
First, for the $s$, $p$ and $d$ waves of the SW system, the absolute value of each parameter decreases with its order for all cases except for $Q_{1,8}$. This behavior guarantees a good convergence at low energies for the ERE, and then, a good description of the ERF and phase-shifts with a small set of effective-range parameters. For the n-p system this effect is rather evident because the ERE converges for all energies.
Second, for the $s$-wave of the SW system, all $Q_{l,n}$ parameters in Table \ref{Texact} have the same sign and almost the same order of magnitude. This fact restricts the range of convergence of the ERE but the numerical estimation for the parameters should be equally good or bad for all, if the numerical precision can handle correctly all calculations of the method, i.e., the results should be reliable except for cases when subtractions or additions of big numbers (from the numerical point of view) are involved in the calculation.
In contrast, this explanation could play a relevant role in the calculation of $Q_{l,n}$ for the n-p system, where the numerical calculation should compute zero from subtractions and additions that maybe involve big numbers.
Third, for the $d$-wave  of SW, the parameters do not have the same sign and the order of magnitude decreases every second order. This increases the range of convergence of the ERE for the $d$-wave.

Let us now show and analyze some results for SW by using the method in Sec.\ \ref{sec-ERE}. For all cases ($s,p,d$-waves) there is only one acceptable value for the channel radius, $a=3$ fm. For $a<3$ fm, part of the potential is neglected and therefore the R-matrix does not give physical results. For $a>3$ fm, the Lagrange mesh technique, and particularly the Gauss quadrature, cannot handle correctly the  discontinuity in the potential at $r=3$ fm and therefore the R-matrix is wrongly estimated. Thus, choosing $a=3$ fm, Table \ref{TSW} displays the absolute error for the first ten effective-range parameters for $s,p,d$-waves of SW. 
\begin{table}[htb]
\caption{Absolute errors of the first ten effective-range parameters for $s,p,d$-waves of SW. The absolute error is written in normalized scientific notation as $\varepsilon=\varepsilon_\text{sig}\times 10^{b}$, where $\varepsilon_\text{sig}$ is the significand and $b$ is the order of magnitude. Here we report $b$ only. Units are shown in Table \ref{Tunits}.}
\label{TSW}
\begin{ruledtabular}
\begin{tabular}{*{10}{>{$}c<{$}}}
	& \multicolumn{3}{c}{$s$-wave} & \multicolumn{3}{c}{$p$-wave} & \multicolumn{3}{c}{$d$-wave}\\
\cline{2-4}\cline{5-7}\cline{8-10}
N			& 10 & 20 & 30 & 10 & 20 & 30 & 10 & 20 & 30 \\
\hline
b(a_l)		& -6  & -9  & -9  & -3  & -6  & -6  & -5  & -8  & -8  \\	
b(r_l)		& -6  & -9  & -9  & -6  & -9  & -9  & -6  & -9  & -9  \\
b(P_l)		& -7  & -12 & -12 & -5  & -9  & -9  & -5  & -9  & -9  \\
b(Q_{l,3})	& -6  & -9  & -9  & -6  & -10 & -10 & -6  & -10 & -10 \\
b(Q_{l,4})	& -6  & -9  & -9  & -6  & -9  & -9  & -6  & -10 & -10 \\
b(Q_{l,5})	& -6  & -9  & -9  & -6  & -10 & -10 & -6  & -10 & -10 \\
b(Q_{l,6})	& -6  & -9  & -9  & -6  & -10 & -9  & -6  & -10 & -10 \\
b(Q_{l,7})	& -6  & -9  & -9  & -7  & -8  & -7  & -7  & -11 & -11 \\
b(Q_{l,8})	& -6  & -9  & -9  & -6  & -8  & -6  & -7  & -11 & -11 \\
b(Q_{l,9})	& -6  & -9  & -9  & -4  & -4  & -4  & -7  & -12 & -12 
\end{tabular}
\end{ruledtabular}
\end{table}

Table \ref{TSW} highlights a good estimation of the effective-range parameters for $N=10$ and very good for $N\ge 20$. This is expected because the square well potential has a polynomial form in the region $r\le 3$ fm, and then, the Lagrange mesh technique and Gauss quadrature give better approximations for the R-matrix than for those potentials with a different dependency on $r$.

A relevant feature in Table \ref{TSW} is the remarkable increasing of $\varepsilon(Q_{1,9})$ in comparison with those of lower order. This effect could be expected because $Q_{1,9}$ is almost two orders of magnitude smaller than the parameters of lower order, which implies that its numerical calculation requires higher computational precision. On the other hand, the ninth order correction in the ERE does not play a relevant role, which can be corroborated by comparing the exact phase-shift and its calculation using the ERE up to eighth and ninth order. This is another way to check whether an effective-range parameter could require higher numerical precision than others of lower order.

Note that in Table \ref{Texact} the absolute values of the effective-range parameters for the $p$-wave of SW look similar to those for the $d$-wave but the estimation is worse for the $p$-wave as Table \ref{TSW} shows. It leads us to think that there is an extra feature that makes poorer estimation for the $p$-wave case (especially for high order parameters). This feature is related with the presence of a weakly bound state, virtual state or resonance at low energy. We shall discuss their effects later in this section.

\begin{figure}[htb]
\includegraphics[width=7.1cm]{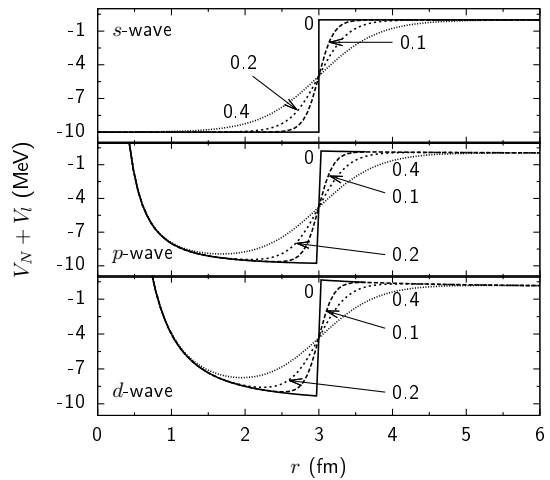}
\caption{Short-range potential $V_N$ plus centrifugal potential $V_l$ for the SW ($d=0$) and W-S systems. Three examples are shown for the W-S system. Labels give the diffuseness in fm.}
\label{Fpot12}
\end{figure}

At this point we have shown that our method in combination with the Lagrange mesh technique works very well for simple potentials. Now we wish to move on to more sophisticated and realistic shapes for the short-range potential. Let us first consider the W-S system. In this case we do not know the exact values of the effective-range parameters to compare with our estimations, therefore in order to have a check point, we shall start with potentials close to the SW potential as Fig.\ \ref{Fpot12} shows. 

The W-S potential does not vanish for a finite distance, which implies that there is not a precise value for $a$. However, the Gauss quadrature should work correctly from the mathematical point of view because this potential has no discontinuity for $d>0$. 

As we discussed at the beginning, the restriction on $a$ is that the short-range potential could be neglected for $r>a$. Thus, in order to obtain acceptable results, we scan the channel radius from a minimal value $\bar{a}$ obtained from $|V_N(\bar{a})|=|V_N(0)/100|$. We also scan the diffuseness in the interval $d\leq 0.6$ fm and choose $N=30$ for all cases.

\begin{figure}[htb]
\includegraphics[width=7.6cm]{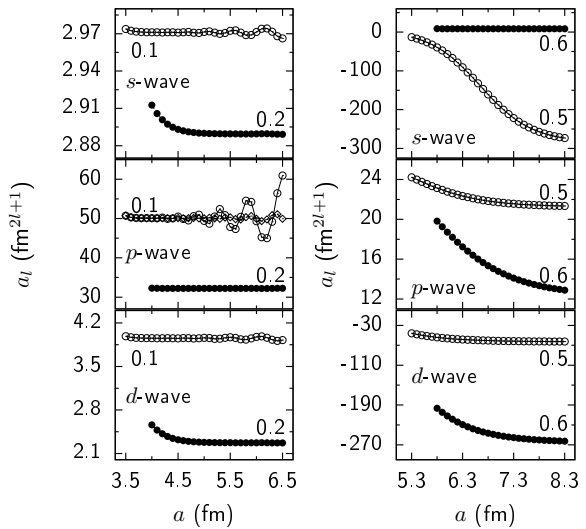}
\caption{Prediction of $a_l$ for the W-S system by choosing different channel radii. For all cases $N=30$ except for the diamonds in the middle-left plot where $N=40$. Labels give the diffuseness in fm. The lines are included to guide the eye.}
\label{Fws}
\end{figure}

In general, we observe a smooth and slow change in the effective-range parameters when the diffuseness increases. Even more, they reach a stable value for channel radii close to $\bar{a}$. An exception to this behavior occurs when a weakly bound state, virtual state or resonance at low energy is present. It is shown in Fig.\ \ref{Fws} where we have chosen the scattering length as an example.

Remember that the choice of the channel radius implies a truncation of the potential range for potentials that do not vanish at a finite distance. This fact in practice, leads to a kind of domino effect, i.e., channel radius $\to$ narrower potential $\to$ slight rise of the energy levels $\to$ chance to modify a weakly bound state making it a virtual state or resonance at low energy $\to$ incorrect estimation of the effective-range parameters. Thus, in order to prevent wrong predictions, the channel radius should be large enough to keep the spectrum unchanged, especially when a weakly bound state is present.

\begin{figure}[htb]
\includegraphics[width=7.6cm]{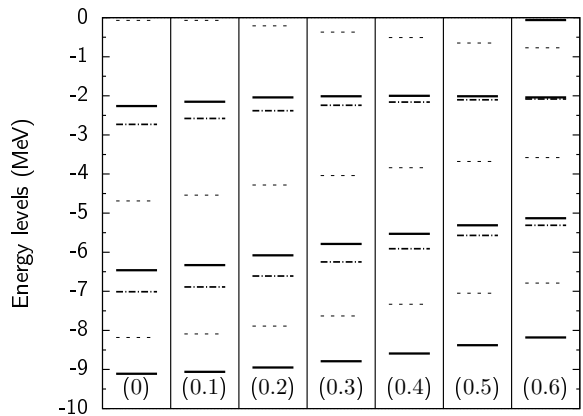}
\caption{Scheme of energy levels for the systems W-S and SW ($d=0$). The diffuseness is shown in parenthesis and is given in fm. \emph{Solid line:} $l=0$. \emph{Dashed line:} $l=1$. \emph{Dashed-dotted line:} $l=2$.}
\label{Fswlevels}
\end{figure}

In order to discuss the impact of virtual states and resonances at low energy and to contrast their effects with those given by weakly bound states, we show the energy levels of the W-S system for different values of $d$ in Fig.\ \ref{Fswlevels}.

From Fig.\ \ref{Fswlevels}, the most evident cases when a weakly bound state is present are $d=0.1$ fm for the $p$-wave and $d=0.6$ fm for the $s$-wave (we skip the case $d=0$ for the $s$-wave, which corresponds to the SW system). Looking at Fig.\ \ref{Fws} for these two cases, one infers that weakly bound states do not introduce new restrictions for the channel radius to those based on the domino effect discussed previously, but may demand a better approximation of the wave function, as it is illustrated by both curves in the case $d=0.1$ and $p$-wave in Fig.\ \ref{Fws}. 

On the other hand, comparing the $s$-wave cross sections at zero energy of the W-S system for $d=0.4$ fm ($\sigma_0(0)\approx 0.1$ b), $d=0.5$ fm ($\sigma_0(0)\approx 10^4$ b) and $d=0.6$ fm ($\sigma_0(0)\approx 10$ b) one sees a remarkable increase of this for the case $d=0.5$ fm. Taking into account that there is no weakly bound state for $d=0.5$ fm as Fig.\ \ref{Fswlevels} shows, then this increase is due to the presence of a virtual state at low energy. The effect of that virtual state on the estimate of $a_0$ is shown on the top-right panel of Fig.\ \ref{Fws}, where the channel radius must be larger than the range that could be chosen by simple inspection of Fig.\ \ref{Fpot12}. The same effect is seen for the $d$-wave and a diffuseness of 0.6 fm. In this case, the $d$-wave cross section shows a resonance around 66 keV which leads to large values of $a$ as the bottom-right panel of Fig.\ \ref{Fws} shows.

Note that for the cases where there is neither a weakly bound state, nor a virtual state or resonance at low energy, Fig.\ \ref{Fws} shows that the estimation of $a_l$ (and in general the effective-range parameters) is more stable. This leads us to summarize the constraints on the channel radius to correctly compute effective-range parameters as: \emph{\textbf{1)} the presence of a weakly bound state can demand a channel radius larger than the typical range of the potential (see for instance the definition of $\bar{a}$) and \textbf{2)} if a virtual state or resonance at low energy is present the channel radius must be much larger}.

We wish to clarify that the previous constraints are based on short-range potentials which are not null but negligible for $r>a$. For potentials strictly vanishing beyond a finite distance, a very good estimation is expected when $a$ is equal to the potential range (see for instance Table \ref{TSW} and Fig.\ \ref{Fswlevels} for the $p$-wave of the SW system).

\begin{figure}[htb]
\includegraphics[width=7.6cm]{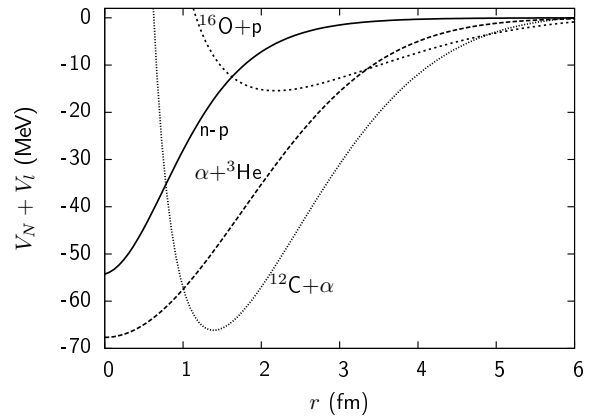}
\caption{Short-range potential $V_N$ plus centrifugal potential $V_l$ for systems n-p ($s$-wave), $\alpha+^3$He ($s$-wave), $^{16}$O$+$p ($p$-wave) and $^{12}$C$+\alpha$ ($d$-wave).}
\label{Fpot3456}
\end{figure}

For the n-p system we also know the exact value of each effective-range parameter. It makes this system a good candidate to test our method. In Fig. \ref{Fpot3456} we show the potential shape for the n-p system and for the three charged cases which will be analyzed later.

\begin{table*}[htb]
\caption{Absolute errors of the first ten effective-range parameters for the n-p system. The absolute error is written in normalized scientific notation as $\varepsilon=\varepsilon_\text{sig}\times 10^{b}$, where $\varepsilon_\text{sig}$ is the significand and $b$ is the order of magnitude. Here we report $b$ only. Units are shown in Table \ref{Tunits}.}
\label{Tn-p}
\begin{ruledtabular}
\begin{tabular}{*{19}{>{$}c<{$}}}
a $ (fm)$	& \multicolumn{3}{c}{8} & \multicolumn{3}{c}{10} & \multicolumn{3}{c}{12} & \multicolumn{3}{c}{15} & \multicolumn{3}{c}{17} & \multicolumn{3}{c}{19} \\
\cline{1-1}\cline{2-4}\cline{5-7}\cline{8-10}\cline{11-13}\cline{14-16}\cline{17-19}
N		& 20 & 30 & 40 & 20 & 30 & 40 & 20 & 30 & 40 & 20 & 30 & 40 & 20 & 30 & 40 & 20 & 30 & 40 \\
\hline
b(a_0)		& -3 & -3 & -3 & -4 & -4 & -4 & -5 & -5 & -5 & -4 & -7 & -7 & -3 & -7 & -9 & -4 & -7 & -9 \\
b(r_0)		& -3 & -3 & -3 & -5 & -5 & -5 & -6 & -6 & -6 & -6 & -8 & -8 & -5 & -9 & -10& -5 & -9 & -9 \\
b(P_0)		& -4 & -4 & -4 & -5 & -5 & -5 & -6 & -6 & -6 & -7 & -7 & -7 & -7 & -9 & -10& -7 & -9 & -8 \\
b(Q_{0,3})	& -2 & -2 & -2 & -2 & -2 & -2 & -3 & -3 & -3 & -5 & -5 & -5 & -6 & -6 & -7 & -6 & -6 & -6 \\
b(Q_{0,4})	& -1 & -1 & -1 & -2 & -2 & -2 & -2 & -2 & -2 & -4 & -4 & -4 & -5 & -5 & -5 & -5 & -5 & -4 \\
b(Q_{0,5})	& -1 & -1 & -1 & -1 & -1 & -1 & -2 & -2 & -2 & -3 & -3 & -3 & -2 & -4 & -2 & -3 & -3 & -3 \\
b(Q_{0,6})	& -1 & -1 & -1 & -1 & -1 & -1 & -1 & -1 & -2 &  0 & -1 & -1 &  0 &  0 &  0 &  0 & 0 &  1 \\
b(Q_{0,7})	&  0 & -1 &  0 &  1 &  1 &  1 &  1 &  2 &  1 &  2 &  2 &  1 &  2 &  2 &  2 &  3 & 3 &  3 \\
b(Q_{0,8})	&  1 &  1 &  2 &  2 &  2 &  3 &  3 &  4 &  4 &  3 &  5 &  5 &  5 &  5 &  5 &  6 & 6 &  5 \\
b(Q_{0,9})	&  3 &  3 &  4 &  5 &  5 &  6 &  6 &  6 &  5 &  7 &  7 &  8 &  8 &  8 &  8 &  8 & 9 &  8 
\end{tabular}
\end{ruledtabular}
\end{table*}
Table \ref{Tn-p} displays the absolute error of the first ten effective-range parameters for the n-p system. Note that, for $N\in\{20,30\}$ and $a\in\{10,12,15\}$ fm, the scattering length, the effective range and the shape parameter are in perfect agreement or slightly better estimated than the results of Ref.\ \cite{PRC63}. The estimation for the first six effective-range parameters is very good if $a=17$ fm and $N=40$. However, the precision for the last four parameters is very bad in general, even increasing the Lagrange mesh points. The possibility of increasing $a$ does not give us better results as one sees in Table \ref{Tn-p} for $a=19$ fm. 

The reader could think that $a=17$ fm is very large following Fig.\ \ref{Fpot3456}, which shows that 4 to 6 fm seems good enough. This increase in the channel radius is expected because the Bargmann potential used for the n-p system provides a virtual state at $-66$ keV. This makes the n-p system a good example where the choice of the channel radius should take into account the presence of a virtual state at low energy.

\begin{figure}[htb]
\includegraphics[width=7.0cm]{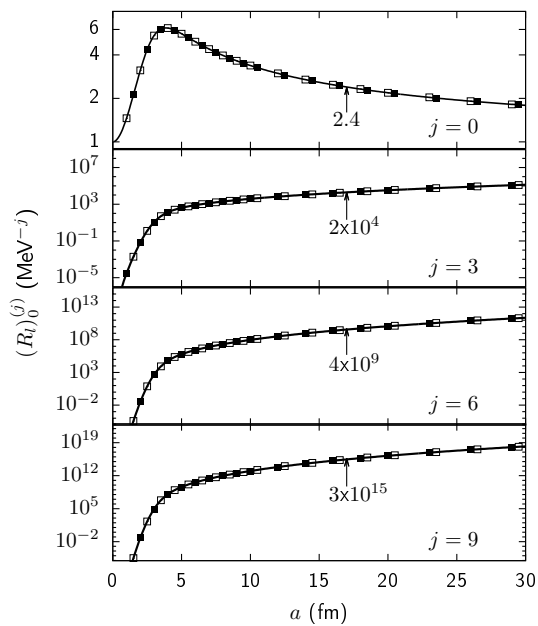}
\caption{R-matrix derivatives at zero energy obtained from the scattering $s$-wave function for the n-p system (solid line) and from numerical calculation of Eq.\ \eqref{dRl0} (square points). Black (white) points correspond to $N=40$ ($N=20$). Four different derivative-orders are shown as examples.}
\label{Fdrmatbargmann}
\end{figure}

The wrong calculation for the last four parameters shown in Table \ref{Tn-p} are explained from the numerical point of view. In those cases, double precision is not enough to keep numbers with a large set of digits, which are needed to compute small quantities from subtractions of big ones. It can be illustrated by comparing intermediate results from the numerical calculation with those obtained analytically as Fig.\ \ref{Fdrmatbargmann} shows for the R-matrix and its derivatives.

In this figure $(R_l)_0^{(j)}$ is computed analytically by using the scattering wave function for a Bargmann potential \cite{Newton,PR107} and numerically by using the Lagrange-mesh technique.
 
Figure \ref{Fdrmatbargmann} highlights that the R-matrix at zero energy ($j=0$) is very well computed numerically for different Lagrange meshes. It agrees with the results for $a_0$ shown in Table \ref{Tn-p}, where one can deduce that the difference between the value of $a_0$ for $N=20$ and $N=40$ is given beyond the fourth digit (remember, the exact value is $a_0=-23.765\cdots$ fm). Similarly for the R-matrix derivatives at zero energy, the bottom three panels of Fig.\ \ref{Fdrmatbargmann} show that the numerical results are in a good agreement with the exact values. 

As the R-matrix derivatives are well computed and the numerical calculation of the $({\mathcal{H}_i}_l)^{(n)}_0$ is accurate, the wrong estimation of the effective-range parameters should come from computational processes. In order to explain it, we wish to emphasize that these parameters are obtained in three steps: First, compute $({\mathcal{H}_i}_l)_0^{(n)}$ and $(R_l)_0^{(j)}$. Second, calculate Eqs.\ \eqref{dDelta}, \eqref{dDeltainv} and \eqref{dDlinv}. Third, determine the effective-range parameters via Eq.\ \eqref{cnn1}. Let us detail the third step by expandig Eq.\ \eqref{cnn1} as
\begin{align}
c_{l,n}=\frac{1}{n!(\Delta_{1l})_0^2}\Biggl[
& \underbrace{{(\mathcal{H}_1}_l)_0^{(n)}{{(\mathcal{H}_2}_l})_0}_{\tau_1}
-\underbrace{{(\mathcal{H}_2}_l)_0^{(n)}{{(\mathcal{H}_1}_l})_0}_{\tau_2}\notag\\
& +\underbrace{a(R_l)^{(n)}_0{(\mathcal{H}'_2}_l)_0({\mathcal{H}_1}_l)_0}_{\tau_3}\notag\\
& -\underbrace{a(R_l)^{(n)}_0({\mathcal{H}'_1}_l)_0({\mathcal{H}_2}_l)_0}_{\tau_4}
+\cdots\Biggr].\label{T}
\end{align}

Equation \eqref{T} shows that numerical limitations can occur if large values of $\tau_1$, $\tau_2$, $\tau_3$ or $\tau_4$ (or any of the remaining terms) appear for a given numerical precision. For instance, by using double precision and choosing $a=17$ fm for the n-p system, one finds that to calculate $c_{0,9}$, which exact value is zero, the terms $\tau_1\approx 2\times 10^{-3}\frac{\text{fm}}{\text{MeV}^9}$, $\tau_2\approx 4\times 10^{-2}\frac{\text{fm}}{\text{MeV}^9}$ and $\tau_3= 0\frac{\text{fm}}{\text{MeV}^9}$ do not contribute to the numerical imprecision as the term $\tau_4\approx 5\times 10^{16}\frac{\text{fm}}{\text{MeV}^9}$ does (the reader can check these values by using Eqs.\ \eqref{mathH1}, \eqref{mathH2} and Fig.\ \ref{Fdrmatbargmann}). Comparing $\tau_3$ and $\tau_4$ one sees that double precision is not enough to compute correctly $c_{0,9}$ and then $Q_{0,9}$. This imprecision together with the fact that $(R_l)^{(n)}_0$ is an approximation (see Eq.\ \eqref{dRl0}) explain the large absolute errors in Table \ref{Tn-p} for parameters of high order. We shall detail the effects of the approximation of $(R_l)^{(n)}_0$ later in the charged case.

As we have illustrated for the n-p system the numerical imprecision comes from big numbers given by R-matrix derivatives (see the previous value of $\tau_4$). This is not a general rule. Sometimes these kind of numbers comes from $({\mathcal{H}_i}_l)^{(n)}_0$ making large values for $\tau_1$ or $\tau_2$. There are three features to get these big numbers for the neutral case (see Eqs.\ \eqref{mathH1} and \eqref{mathH2}): \textbf{1)} large channel radius, \textbf{2)} big reduced mass and \textbf{3)} high order of the effective-range parameter.

When numerical inaccuracies appear for the effective-range parameters of high order, one can give a gross approximation of these parameters. Let us illustrate it using Table \ref{Tn-p}. For the last four effective-range parameters the imprecision decreases for small channel radii. It is totally expected following the argument number 1 presented previously. On the other hand, the precision for the first six parameters decreases for small channel radii (e.g., $N\in\{30,40\}$ and $a\leq 17$ fm), which is due to the approximation for the potential at $r>a$ and not to numerical limitations. These two behaviors allow us to predict the order of magnitude for the first nine or ten effective-range parameters. It is achieved by choosing a small channel radius to avoid numerical inaccuracies for parameters of high order, but not too small to keep the correct order of magnitude for those of low orders. If an effective-range parameter is zero its order of magnitude is indeterminate, and therefore, we suggest to report the absolute error (e.g., for $a=5$ fm and $N=40$, the absolute error falls in the range $[0.007,0.7]\text{ fm}^{2n-1}$ for the first nine effective-range parameters of the n-p system).

\subsection{Charged cases}\label{sec-app-charged}
Comparing Fig.\ \ref{Fpot12} and Fig.\ \ref{Fpot3456}, one expects that the channel radius for the $^{3}$He$+\alpha$, $^{16}$O$+$p and $^{12}$C$+\alpha$ systems will be larger than the channel radius of the W-S system. This intuitive prediction has been corroborated in Ref.\ \cite{PRC63} for the $^{3}$He$+\alpha$ and $^{16}$O$+$p systems, where varying the channel radius from 10 fm up to 14 fm, the scattering lengths, the effective ranges and the shape parameters are stable. Therefore, starting with the values of $N$ and $a$ which give us the best prediction according to Ref.\ \cite{PRC63}, and using the method described in Sec.\ \ref{sec-met-charged}, we display in Table \ref{Tcharged1} the numerical results for the first ten effective-range parameters of the systems $^{3}$He$+\alpha$ and $^{16}$O$+$p. 
\begin{widetext}
\vspace*{-1cm}
\begin{center}
\begin{table}[htb]
\caption{Results for the first ten effective-range parameters for $^{3}$He$+\alpha$ and $^{16}$O$+$p. Units are shown in Table \ref{Tunits}.}
\label{Tcharged1}
\begin{ruledtabular}
\begin{tabular}{*{19}{>{$}c<{$}}}
	& \multicolumn{4}{c}{$^{3}$He$+\alpha$ ($s$-wave)} & \multicolumn{4}{c}{$^{16}$O$+$p ($p$-wave)}  \\
\cline{2-5}\cline{6-9}
a $ (fm)$ & \multicolumn{2}{c}{12} & \multicolumn{2}{c}{14} & \multicolumn{2}{c}{14} & \multicolumn{2}{c}{16} \\
\cline{1-1}\cline{2-3}\cline{4-5}\cline{6-7}\cline{8-9}
N		& 30 & 40 & 30 & 40 & 30 & 40 & 30 & 40 \\
\hline
a_l		& 36.8849 				& 36.8863 				& 36.8847 			& 36.8862 			& 401.99 				&  401.91				& 401.77			& 401.89			\\
r_l		& 0.97262 				& 0.97263 				& 0.97262 			& 0.97262 			& -0.02909 				& -0.02910			& -0.02912			& -0.02910			\\
P_l		& -0.09010 				& -0.09009 				& -0.09009 			& -0.09008 			& 10233 				&  10217				& 10193				& 10212				\\
Q_{l,3}	& 0.04972 				& 0.04970 				& 0.04972 			& 0.04971 			& -0.75049 				& -0.750595			& -0.75044			& -0.75045			\\
Q_{l,4}	& 0.09278 				& 0.09281 				& 0.09281 			& 0.09282 			& -0.69557 				& -0.69538			& -0.69557			& -0.69619			\\
Q_{l,5}	& -0.07573 				& -0.07571 				& -0.07581 			& -0.075771 		& 4.83493 				&  4.836839			& 4.83510			& 4.83321			\\
Q_{l,6}	& -3.12\times 10^{-1}  	& -3.12\times 10^{-1}	& -7.45\times 10^{-2}	& -7.46\times 10^{-2} 	& 14.8754 				&  14.8789				& 14.6309			& 14.6329			\\
Q_{l,7}	& 6.92\times 10^1 		& 6.92\times 10^1 		& -4.93\times 10^2 	& -4.93\times 10^2 	& 2.37\times 10^2 		&  2.37\times 10^2		& 4.61\times 10^2	& 4.61\times 10^2	\\
Q_{l,8}	& 6.21\times 10^6 		& 6.21\times 10^6 		& 3.39\times 10^6 	& 3.39\times 10^6 	& 4.67\times 10^6 		&  4.67\times 10^6		& -1.20\times 10^6	& -1.19\times 10^6	\\
Q_{l,9}	& -1.65\times 10^{10} 	& -1.65\times 10^{10} 	& -3.41\times 10^8 	& -3.41\times 10^8 	& -2.33\times 10^{10} 	& -2.33\times 10^{10}	& 2.47\times 10^9	& 2.47\times 10^9	
\end{tabular}
\end{ruledtabular}
\end{table}
\end{center}
\vspace*{-1cm}
\end{widetext}

For both systems, the estimation of $a_l$, $r_l$ and $P_l$ are in good agreement with the results of Ref.\ \cite{PRC63}. The next three parameter are stable and their values agree with the description of the phase-shifts at low energies, which can be checked by using Eqs.\ \eqref{psl2} and \eqref{Kl}. These facts indicate that $Q_{l,i}$ is correctly estimated for $i<6$. The last three parameters are evidently greater than those of lower order, which indicates that numerical limitations are present by using double precision calculations.

The $^{3}$He$+\alpha$ and $^{16}$O$+$p systems do not have weakly bound states (our potential models provide bound states at $-34.2$ MeV and $-5.7$ MeV for $^{3}$He$+\alpha$ and $-3.6$ MeV for $^{16}$O$+$p) and the phase-shits at low energies do not show indication about possible resonances. It is coherent with the stability of $a_l$ in Table \ref{Tcharged1}, and corroborates that the imprecisions come from the numerical limitations.

In order to detail the numerical limitations, we should track where the large numbers come from in the numerical calculation. To do that we need to evaluate the contribution of the R-matrix derivatives and the renormalized Coulomb functions, in a similar way as we have shown in Eq.\ \eqref{T}. Note that this equation is not valid for the charged case, but removing the term $n!$ from the denominator on the right side, one gets $(D_l^{-1})^{(n)}_0$ (see Eq.\ \eqref{cnn1}), which has the same structure for the neutral and charged case. Thus, we can compute the terms $\tau_1$, $\tau_2$, $\tau_3$ and $\tau_4$ and see if numerical imprecisions appear for determining $(D_l^{-1})^{(9)}_0$.

For the $^{3}$He$+\alpha$ system one finds ($a=14$ fm, $N=40$)
{\centering
$\tau_1\approx -0.2 \text{ fm/MeV}^9, \quad \tau_2\approx 5\times 10^{11} \text{ fm/MeV}^9$,

$\tau_3\approx -5\times 10^6 \text{ fm/MeV}^9 \quad\text{and}\quad\tau_4\approx 6\times 10^6 \text{ fm/MeV}^9$.}

For the $^{16}$O$+$p system one finds ($a=16$ fm, $N=40$)
{\centering
$\tau_1\approx 10^3 \text{ fm/MeV}^9, \quad
\tau_2\approx 4\times 10^8 \text{ fm/MeV}^9$, 

$\tau_3\approx -7\times 10^3 \text{ fm/MeV}^9 \quad\text{and}\quad
\tau_4\approx 9\times 10^3 \text{ fm/MeV}^9.$}

\begin{figure}[htb]
\includegraphics[width=7.6cm]{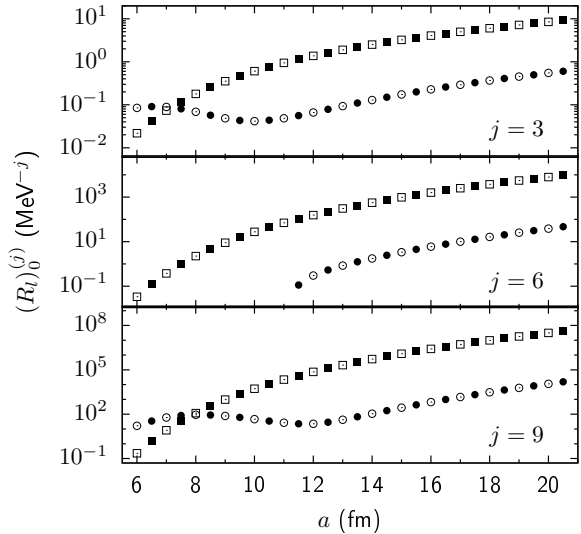}
\caption{Numerical results of $(R_l)_0^{(j)}$ for $^{3}$He$+\alpha$ (square points) and $^{16}$O$+$p (circle points). Black (white) points correspond to $N=40$ ($N=30$).}
\label{Fdrmat4-5}
\end{figure}

In Fig.\ \ref{Fdrmat4-5} we present numerical results of $(R_l)_0^{(j)}$ for $j\in\{3,6,9\}$. This figure shows that the contribution of $(R_l)_0^{(j)}$ to form big numbers increases with $j$, as one could expect. However, despite rather large values of the R-matrix derivatives at zero energy ($(R_l)_0^{(j)}\lesssim 10^7$ MeV$^{-j}$ if $j\leq 9$ and $a\leq 16$ fm for both systems), the values of $\tau_3$ and $\tau_4$ indicate that there is not a relevant influence of these derivatives for numerical inaccuracies. Of course, this argument is valid if $(R_l)_0^{(j)}$ is computed precisely (we shall discuss it later). 

In contrast, Table \ref{TCoulfunc456} shows that the values of the renormalized Coulomb functions and their derivatives can be large enough to create numerical inaccuracies as seen by the large values of $\tau_2$ with respect to those of $\tau_3$ and $\tau_4$.
\begin{table}[htb]
\caption{Values for the renormalized Coulomb functions and some of their derivatives at zero energy for $^{3}$He$+\alpha$, $^{16}$O$+$p and $^{12}$C$+\alpha$. ${\mathcal{H}_i}_l^{(n)}$ is given in fm$^{1/2}/$MeV$^n$.}
\label{TCoulfunc456}
\begin{ruledtabular}
\begin{tabular}{>{$}c<{$}>{$}c<{$}>{$}c<{$}>{$}c<{$}}
                 & ^{3}$He$+\alpha	& ^{16}$O$+$p$ & ^{12}$C$+\alpha  \\
                 & (s$-wave$)  & (p$-wave$) & (d$-wave$)  \\
a \text{ (fm)}   & 14             	& 16           & 17   \\
a_N \text{ (fm)} & 4.22           	& 3.83         & 0.81 \\
\hline
({\mathcal{H}_1}_l)_0^{(0)} & 186             & 170            &  1.3\times 10^5    \\
({\mathcal{H}_2}_l)_0^{(0)} & 0.02            & 0.02           &  1.4\times 10^{-5} \\
({\mathcal{H}_1}_l)_0^{(3)} & -299            & 108            & -1.8\times 10^5    \\
({\mathcal{H}_2}_l)_0^{(3)} & 1.68            & 0.1            &  1.6\times 10^{-4} \\
({\mathcal{H}_1}_l)_0^{(6)} & 83.6            & -1016          & -5.8\times 10^5    \\
({\mathcal{H}_2}_l)_0^{(6)} & 9698            & 81.6           &  1.1\times 10^{-2} \\
({\mathcal{H}_1}_l)_0^{(9)} & -7.4            & 6.6\times 10^4 &  5.9\times 10^6    \\
({\mathcal{H}_2}_l)_0^{(9)} & 2.8 \times 10^9 & 2.1\times 10^6 &  4.9               \\
\end{tabular}
\end{ruledtabular}
\end{table}
Note that the results for $\tau_1$, $\tau_2$, $\tau_3$ and $\tau_4$ can be obtained partially by using Fig.\ \ref{Fdrmat4-5} and Table \ref{TCoulfunc456} (the numbers in Table \ref{TCoulfunc456} are rounded).

Following Table \ref{Tcharged1}, numerical inaccuracies appear calculating $Q_{0,6}$ for the $^{3}$He$+\alpha$ system and calculating $Q_{1,7}$ for the $^{16}$O$+$p system, i.e., to higher order for the latter. It is not surprising from the numerical point of view because the numbers involved in the calculation of the effective-range parameters are smaller for the $^{16}$O$+$p system. We think that this could be part of the explanation about why the parameter $Q_{1,6}$ seems more stable than the parameter $Q_{0,6}$ in Table \ref{Tcharged1}.

It is clear that the numerical inaccuracies are stronger for the $^{3}$He$+\alpha$ system than for the $^{16}$O$+$p system, and if the R-matrix derivatives are well computed, it is also clear that the most significant contribution to make big numbers comes from the renormalized Coulomb functions for both systems. Let us suppose for a moment that the numerical calculations were performed by using single precision. In this case, single precision cannot handle correctly additions and subtractions of $\tau_1$, $\tau_2$, $\tau_3$ and $\tau_4$ for the $^{3}$He$+\alpha$ and $^{16}$O$+$p systems, and then we could stop the discussion at this point. However, all results shown here are obtained by using double precision, which can handle correctly those additions and subtractions. Therefore, the only possibility to get numerical imprecision comes from the approximation of $(R_l)_0^{(j)}$, which can also  be understood as a numerical limitation. 

As Eq.\ \eqref{dRl0} stresses, a better estimation of the R-matrix derivatives at zero energy is achieved by increasing the Lagrange mesh points, $N$. It means that there exists a value of $N$, $\bar{N}$, for which one gets a \csi stable" approximation of $(R_l)_0^{(j)}$. Here, \csi stable"  means that for $N\geq \bar{N}$ the value of $(R_l)_0^{(j)}$ does not change up to a specific digit in its fractional part. 

Let us illustrate the role that the previous effect plays for the $^{16}$O$+$p system. Figure \ref{Fdrmat4-5} shows that there is not a significant difference in the estimation of $(R_l)_0^{(j)}$ by using $N=30$ or $N=40$. For these Lagrange meshes and $a=16$ fm, we obtain $(R_1)_0^{(9)}=656.590\cdots$ MeV$^{-9}$ and $(R_1)_0^{(9)}=656.565\cdots$ MeV$^{-9}$ respectively, which means that we have a precise determination of $(R_1)_0^{(9)}$ up to the first digit in the fractional part. Our internal calculations have shown that a precise determination of the second digit in the fractional part of $(R_1)_0^{(9)}$ is possible for $N>120$. This effect makes a clear numerical limitation to determine the effective-range parameters by using the technique in Sec.\ \ref{sec-LM}, where in order to compute correctly a large set of digits for $(R_l)_0^{(j)}$, one should implement high-precision in the numerical calculation because of the large mesh required. 

It is important to stress that the previous numerical limitation is remarkably strong when the value of $(R_l)_0^{(j)}$ are large, which is often for determining effective-range parameters of high order (see n-p system as an example).

On the other hand, in the charged case there is an extra parameter, which is the nuclear Bohr radius $a_N$ or the nuclear Rydberg energy $E_N$. Remember that $E_N$ can be written in terms of $a_N$ (see Eq.\ \eqref{eta}). Hence, in order to explore the impact of $a_N$, we have introduced the $^{12}\text{C}+\alpha$ system, which has a nuclear Bohr radius five times smaller (approximately) than the  nuclear Bohr radius of $^{3}$He$+\alpha$ or $^{16}$O$+$p. 

Table \ref{Tcharged2} shows that the first ten effective-range parameters associated to $^{12}\text{C}+\alpha$ are stable, which agrees with the results presented in Table \ref{TCoulfunc456} and the description of the phase shifts following Eqs.\ \eqref{psl2} and \eqref{Kl} at low energies. This stability and description are also expected from Eqs. \eqref{dErH1serie2} and \eqref{dErH2serie2}, where one sees that derivatives of the renormalized Coulomb functions are inversely proportional to some power of the nuclear Rydberg energy (see for instance Eq.\ \eqref{dsqrtwl}, which is necessary to compute Eqs. \eqref{dErH1serie2} and \eqref{dErH2serie2}) or directly proportional to some power of the nuclear Bohr radius. It leads us to infer that the numerical limitations introduced by the renormalized Coulomb functions will be more important when $a_N$ is large or when $E_N$ is low.
\begin{table}[htb]
\caption{Results for the first ten parameters in the ERE for $^{12}$C$+\alpha$. Units are shown in Table \ref{Tunits}.}
\label{Tcharged2}
\begin{ruledtabular}
\begin{tabular}{*{5}{>{$}c<{$}}}
a $ (fm)$ & \multicolumn{2}{c}{15} & \multicolumn{2}{c}{17} \\
\cline{1-1}\cline{2-3}\cline{4-5}
N		& 30 & 40 & 30 & 40 \\
\hline
a_2		& 58926		& 58903		& 58899		& 58903		\\
r_2		& 0.15797	& 0.15797	& 0.15797	& 0.15797	\\
P_2		& -65.964	& -65.964	& -65.964	& -65.964	\\
Q_{2,3}	& 0.14105	& 0.14105	& 0.14105	& 0.14105	\\
Q_{2,4}	& -0.00986	& -0.00986	& -0.00987	& -0.00986	\\
Q_{2,5}	& -0.02996	& -0.02997	& -0.02997	& -0.02997	\\
Q_{2,6}	& -0.02971	& -0.02973	& -0.02975	& -0.02973	\\
Q_{2,7}	& -0.00154	& -0.00155	& -0.00157	& -0.00158	\\
Q_{2,8}	& 0.05553	& 0.05557	& 0.05550	& 0.05547	\\
Q_{2,9}	& 0.13318	& 0.13393	& 0.13035	& 0.12977	
\end{tabular}
\end{ruledtabular}
\end{table}

Finally, we wish emphasize that the R-matrix and its derivatives for $^{12}$C$+\alpha$ do not have large values to introduce strong numerical inaccuracies. For instance, for $a=17$ fm and $N=40$, $(R_l)_0^{(3)}\approx 0.1$ MeV$^{-3}$, $(R_l)_0^{(6)}\approx -5$ MeV$^{-6}$ and $(R_l)_0^{(9)}\approx 3\times 10^5$ MeV$^{-9}$. This means that for computing $(D_l^{-1})^{(9)}_0$ there will be numbers such as $\tau_1\approx 85\frac{\text{fm}}{\text{MeV}^9}$, $\tau_2\approx 6\times 10^5\frac{\text{fm}}{\text{MeV}^9}$, $\tau_3\approx -4\times 10^6\frac{\text{fm}}{\text{MeV}^9}$ and $\tau_4\approx 4\times 10^6\frac{\text{fm}}{\text{MeV}^9}$, which can be checked partially, by using Table \ref{TCoulfunc456} and $(R_l)_0^{(9)}$.

\section{Conclusions and perspectives}\label{sec-CP}
Calculations of the effective range parameters for neutral and charged nuclear systems are accurately carried out  by solving the Schrödinger equation at zero energy for a given interaction potential, using the R-matrix and Lagrange-mesh methods. The expressions developed here show an easy path to go beyond the second order in the ERE without any mathematical or physical limitation. In particular, the present results are much more rigorous and systematic than methods based on a fit of low-energy scattering phase shifts, which typically lead to large error bars for high-order parameters \cite{PRC81}.

Nevertheless, problems in the numerical calculation of the present method can arise due to computational imprecisions, in particular for large values of the channel radius $a$. In general, choosing the minimal $a$ such that the short range potential is negligible at distances larger than $a$, leads to a satisfactory gross estimate of the effective-range parameters. However, to improve the accuracy on the lowest-order parameters, larger radii are needed which lead to numerical instabilities for the high-order parameters. Using double numerical precision, the expressions and methods explained here have worked very well up to the fifth or sixth order in energy on all examples considered above. In some cases, a high accuracy could even be reached at least up to the ninth order. 

For typical meshes (30 to 40 mesh points) the Lagrange-mesh technique gives a fast convergence for the first digits of the R-matrix derivatives at zero energy. However, the present method requires a large set of digits for these derivatives in order to compute the effective-range parameters precisely. This implies a much larger mesh which in general demands an implementation with high numerical precision. If such implementation is made, the numerical limitation given by large channel radius should be also decreased.

When a weakly bound state is present a large channel radius can be demanded in order to keep the physical meaning of this state and the effective-range parameters. Similarly, if there is a virtual or resonant state at energies very close to the threshold, choosing a large channel radius is mandatory in order to estimate the effective range parameters with a good precision. This case can thus be particularly delicate to handle since large radii can lead to numerical instabilities. However, when applied to charged systems, our results have shown that the Coulomb barrier partly solves this problem. We infer that in general high-accuracy predictions of the effective-range parameters can be obtained without numerical difficulty for nuclear systems with a large nuclear Rydberg energy or a small nuclear Bohr radius. The contribution of the centrifugal barrier could also play a secondary role in this improvement and would deserve a further study.

The present work thus opens a window to get a better description of elastic scattering for systems where the physical interest is at the low-energy regime, for instance in nuclear astrophysics. With this new tool, the problem of the connection between low-energy scattering phase shifts and weakly-bound-state asymptotic normalization constants (ANCs) could be revisited, at least for given interaction potentials. This is for instance important in systems of astrophysical interest like $^{12}$C$+\alpha$ \cite{PRC81}. Once this connection is well understood for theoretical models, the possibility to directly extract ANCs from experimental data could be considered.

Another perspective is the use of Padé expansions rather than Maclaurin expansions for the effective-range function. These expansions are expected to be valid on a wider energy range and to allow the description of resonances, two useful features for a direct fit of experimental data \cite{JPCS312}. These expected features could be checked on theoretical models first, hopefully leading to general prescriptions to prefer one type of expansion to another.

\section*{Acknowledgments}
We warmly thank to E. C. Pinilla and D. Baye for valuable discussions and suggestions. This text presents research results of the IAP program P7/12 initiated by the Belgian-state Federal Services for Scientific, Technical, and Cultural Affairs. O.L.R.S. is supported by the IAP program.

\end{document}